\documentclass[a4paper,fleqn,usenatbib]{mnras}

\usepackage{newtxtext,newtxmath,hyperref}

\usepackage[T1]{fontenc}
\usepackage{ae,aecompl}

\usepackage{graphicx}	
\usepackage{amsmath}	
\usepackage{amssymb}	

\newcommand{\Msun}{\ensuremath{\mathrm{M}_\odot}}
\newcommand{\ME}{\ensuremath{m_{\rm{e}}}}
\newcommand{\MP}{\ensuremath{m_{\rm{p}}}}

\newcommand{\KB}{\ensuremath{k_{\rm{B}}}}
\newcommand{\HI}{\ensuremath{\mathrm{H}_{\mathrm{I}}\,}}
\newcommand{\HII}{\ensuremath{\mathrm{H}_{\mathrm{I\hspace{-.1em}I}}\,}}

\title[X-ray Novae produced by Isolated BHs]{Can Isolated Single Black Holes Produce X-ray Novae?}

\author[T. Matsumoto et al.]{
Tatsuya Matsumoto,$^{1}$\thanks{E-mail: matsumoto@tap.scphys.kyoto-u.ac.jp}
Yuto Teraki,$^{2,3}$
and Kunihito Ioka$^{2}$
\\
$^{1}$Department of Physics, Graduate School of Science, Kyoto University, Kyoto 606-8502, Japan\\
$^{2}$Center for Gravitational Physics, Yukawa Institute for Theoretical Physics, Kyoto University, Kyoto 606-8502, Japan\\
$^{3}$National Institute of Technology, Asahikawa College, Asahikawa 071-8142, Japan
}

\date{Accepted XXX. Received YYY; in original form ZZZ}

\pubyear{2017}

\begin{document}
\label{firstpage}
\pagerange{\pageref{firstpage}--\pageref{lastpage}}
\maketitle

\begin{abstract}
Almost all black holes (BHs) and BH candidates in our Galaxy have been discovered as soft X-ray transients, so-called X-ray novae.
X-ray novae are usually considered to arise from binary systems.
Here we propose that X-ray novae are also caused by isolated single BHs.
We calculate the distribution of the accretion rate from interstellar matter to isolated BHs, and find that BHs in molecular clouds satisfy the condition of the hydrogen-ionization disk instability, which results in X-ray novae.
The estimated event rate is consistent with the observed one.
We also check an X-ray novae catalog and find that $16/59\sim0.27$ of the observed X-ray novae are potentially powered by isolated BHs.
The possible candidates include \textit{IGR J17454-2919}, \textit{XTE J1908-094}, and \textit{SAX J1711.6-3808}.
Near infrared photometric and spectroscopic follow-ups can exclude companion stars for a BH census in our Galaxy.
\end{abstract}




\section{Introduction}\label{Introduction}
X-ray novae are soft X-ray transient events with $\sim10$ days of rapid brightening up to $\sim10^{38}\,\rm{erg\,s^{-1}}$, followed by exponential decays \citep[see][for reviews]{1996ARA&A..34..607T,1997ApJ...491..312C,2015ApJ...805...87Y}.
X-ray novae are considered to be produced by binary systems composed of low mass stars and compact objects such as neutron stars or black holes (BHs), so-called a low mass X-ray binary (LMXB) system.
About twenty LMXBs have been dynamically confirmed to contain BHs through spectral observations of companion stars.
Furthermore, 30-40 of X-ray novae share the same X-ray signatures with those of BH LMXB systems, but they are too faint after an outburst to conduct follow-up observations \citep{2016A&A...587A..61C}.
Therefore, strictly speaking, it is unclear whether these X-ray novae without follow-up observations are really produced by binary systems or not.

\begin{figure}
\includegraphics[width=0.7\columnwidth,angle=270]{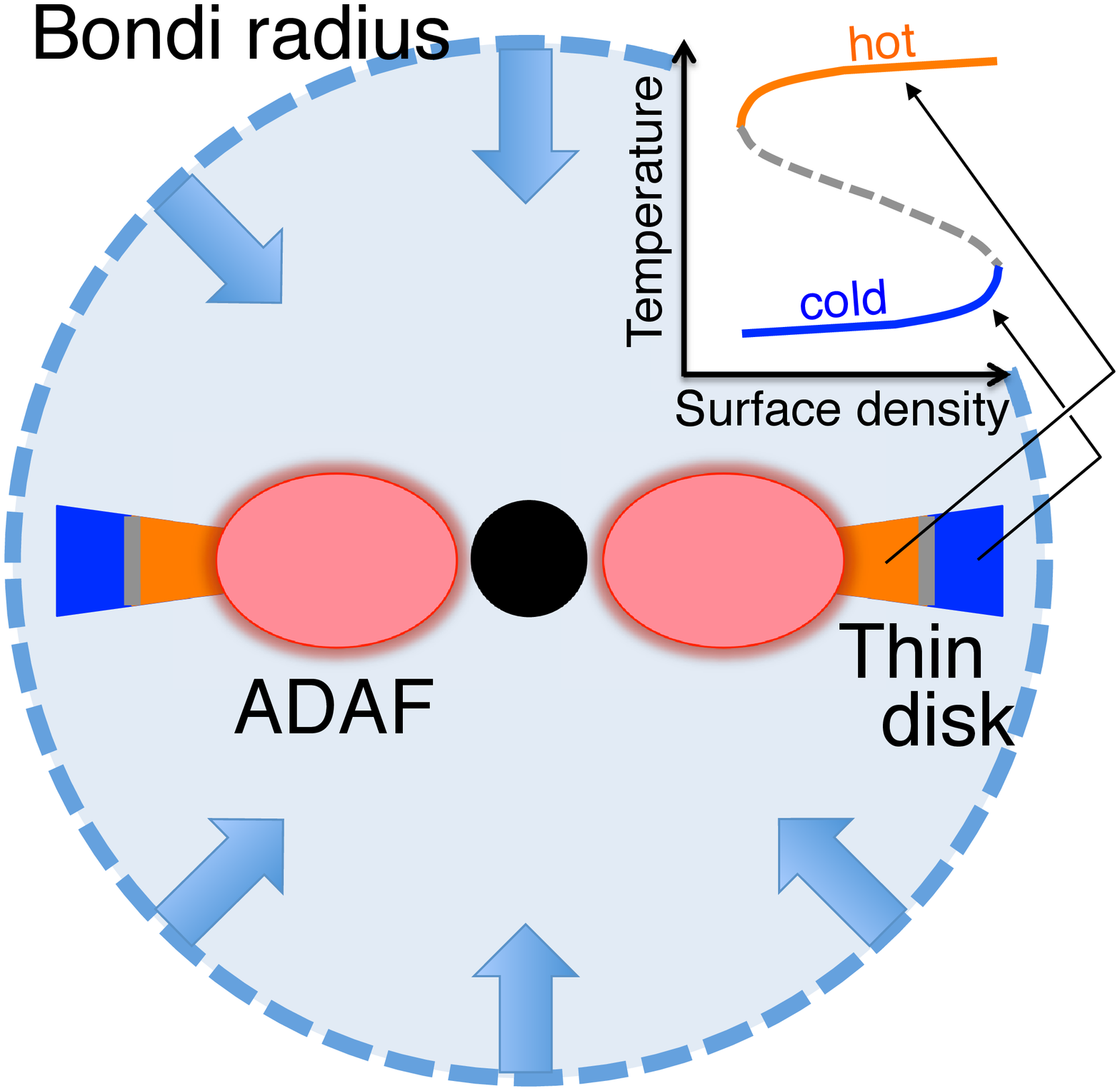}
\caption{Schematic picture of an isolated BH with a thin disk part attached to inner ADAF part. The thin disk part consists of an outer cold branch and an inner hot branch in the S-shaped thermal equilibrium curve (upper right).}
\label{fig pic}
\end{figure}

In the standard scenario, X-ray novae are explained by the hydrogen-ionization instability of accretion disks (a kind of thermal-viscous instability, see \cite{2001NewAR..45..449L}, for a review).
The instability model is originally proposed as a mechanism of dwarf novae, which are optical transients caused by white dwarf binary systems \citep{1974PASJ...26..429O,1979PThPh..61.1307H}, and later applied to X-ray novae \citep{1984AIPC..115...49V,1985ASSL..113..307C,1989ApJ...343..229H,1989ApJ...343..241M}.
When a region in a disk has low temperature for hydrogen to recombine, the negative hydrogen ($\rm{H^-}$) ions dominate the opacity.
As opposed to the free-free opacity, the $\rm{H^-}$ hydrogen opacity is a steep increasing function of temperature.
This rapid response to the temperature change makes an S-shaped structure in the thermal equilibrium curve of the region (see Fig \ref{fig pic}).
In the quiescent phase, the region is in the low temperature branch, while when the surface density reaches a critical value, the region makes a transition to a hot branch and increases the mass accretion rate.
Then, the inner annulus is also heated, and the whole disk mass accretes to the central object.
This is the origin of X-ray brightening.

Recently, the advanced Laser Interferometer Gravitational Observatory (LIGO) has detected gravitational waves (GWs) and observed binary BH mergers for the first time \citep{2016PhRvL.116f1102A,2016PhRvL.116x1103A,2016PhRvX...6d1015A}.
If such merged spinning BHs exist in our Galaxy, they can be high energy sources \citep{2017MNRAS.470.3332I}. 
Before the GW detections, isolated BHs are also believed to reside in our Galaxy.
Based on the stellar evolution theory, the number of isolated BHs is as many as $\sim10^8$ \citep{1983bhwd.book.....S}, some authors have discussed high energy phenomena caused by isolated BHs \citep{1999ApJ...523L...7A,2012MNRAS.427..589B,Teraki}, and studied the detectability of isolated BHs \citep{1998ApJ...495L..85F,2002MNRAS.334..553A,2013MNRAS.430.1538F,Matsumoto}.
However, because of the very low mass accretion rate, the accretion disks around BHs are radiatively inefficient \citep{1994ApJ...428L..13N} and the  detection of isolated BHs is challenging if BHs are stationary sources.  

In this paper, we propose a novel idea that isolated single BHs in our Galaxy can produce transient events like X-ray novae.
The structure of this paper is as follows.
First, we calculate the mass accretion distribution of isolated BHs (section \ref{Accretion Rate Distribution of isolated BHs}).
Next, we find that some BHs in molecular clouds accrete enough mass to have a thin disk part in section \ref{X-ray Novae produced by Isolated BHs}.
This is because for a large mass accretion rate, a disk has a high density enough to cool by radiation and become geometrically thin at the outer part of the disk \citep{1995ApJ...438L..37A,1995ApJ...452..710N}.
In Fig \ref{fig pic}, we show a schematic picture of the system we consider.
The thin disk region can suffer from the hydrogen-ionization instability and cause transient events like X-ray novae.
We also estimate that the event rate is comparable to that of the observed X-ray novae.
Finally, we suggest that some X-ray novae without companions could be produced by isolated BHs in section \ref{Discussion}.
We also discuss how to discriminate an isolated BH from a binary by observations.

\section{Accretion Rate Distribution of isolated BHs}\label{Accretion Rate Distribution of isolated BHs}
In this section, we consider the mass accretion onto Galactic isolated BHs.
The BHs accrete interstellar medium (ISM) gas and form accretion disks.
We study the accretion rate distribution of isolated BHs taking the BH mass, the BH velocity, and the ISM density distributions into account.

An isolated BH pulls the ISM gas by its gravity and accretes the gas via so-called Bondi accretion \citep{1939PCPS...35..405H,1944MNRAS.104..273B,1952MNRAS.112..195B}.
The Bondi radius and the accretion rate are estimated as
\begin{eqnarray}
R_{\rm{B}}&=&\frac{GM}{c_{\rm{s}}^2+v^2}
   \label{bondi radius}\\
&\simeq&8.3\times10^{13}\,\biggl(\frac{M}{10\,\Msun}\biggl)\biggl(\frac{V}{40\,\rm{km\,s^{-1}}}\biggl)^{-2}{\,\rm{cm}},\\
\dot{M}_{\rm{B}}&=&4\pi{R_{\rm{B}}}^2\rho{V}
   \label{bondi rate}\\
&\simeq&8.2\times10^{11}\,\biggl(\frac{M}{10\,\Msun}\biggl)^{2}\biggl(\frac{V}{40\,\rm{km\,s^{-1}}}\biggl)^{-3}\biggl(\frac{n}{1\,\rm{cm^{-3}}}\biggl){\,\rm{g\,s^{-1}}},
   \label{bondi rate2}
\end{eqnarray}
where, $G$, $M$, $v$, $c_{\rm{s}}$, and $\rho$ are the gravitational constant, the BH mass, the BH velocity, the sound speed of ISM, and the mass density of ISM, respectively.
We define the velocity $V$ as $V:=\sqrt{c_{\rm{s}}^2+v^2}$.
From Eq. \eqref{bondi rate} to Eq. \eqref{bondi rate2}, we convert the mass density to the number density $n$ by using the relation $\rho=m_{\rm{u}}\mu{n}$, where $m_{\rm{u}}$ and $\mu=1.41$ are the atomic mass unit and the mean molecular weight.
It should be noted that the Bondi accretion rate is much smaller than the Eddington accretion rate defined as 
\begin{eqnarray}
\dot{M}_{\rm{Edd}}&=&\frac{4\pi{GMm_{\rm{p}}}}{\eta\sigma_{\rm{T}}c}
 \label{eddington rate}\\
&\simeq&1.4\times10^{19}\,\biggl(\frac{M}{10\,\Msun}\biggl){\,\rm{g\,s^{-1}}},
\end{eqnarray}
where $m_{\rm{p}}$, $\eta=0.1$, $\sigma_{\rm{T}}$ and $c$ are the proton mass, radiative efficiency\footnote{Some literatures define the Eddington accretion rate without the radiative efficiency $\eta$. It should also be noted that the name ``radiative efficiency" does not mean the true efficiency.}, Thomson cross section, and the speed of light, respectively.

The accreting matter forms an accretion disk at the centrifugal radius \citep{1976ApJ...204..555S}.
In the Bondi accretion, the accreting gas is assumed to have a spherical or axial symmetry.
However, the actual ISM has density and turbulent velocity fluctuations.
Because of the difference of the density and the turbulent velocity in the scale of the Bondi radius, the accreting ISM has a specific angular momentum of
\begin{eqnarray}
l=\frac{1}{4}\biggl[\frac{\delta{\rho}}{\rho}\if{\biggl|_{R=2R_{\rm{B}}}}\fi+2\frac{\delta{V}}{V}\if{\biggl|_{R=2R_{\rm{B}}}}\fi\biggl]R_{\rm{B}}V.
\end{eqnarray}
The quantities $\delta{\rho}/\rho$ and $\delta{V}/V$ are evaluated at the Bondi radius length scale ($\sim2R_{\rm{B}}$).
For the former quantity, we use the observed density fluctuation as $\delta{\rho}/\rho\simeq(R/6\times10^{18}\,\rm{cm})^{1/3}$ \citep{1995ApJ...443..209A,2011piim.book.....D}.
This scaling holds at smaller length scale than the Bondi radius.
The latter quantity is evaluated by using the correlation of the turbulent velocity dispersion and the length scale in the molecular clouds, so-called Larson's first law as \citep{1981MNRAS.194..809L,2004ApJ...615L..45H}, $\delta{v_{\rm{turb}}}\simeq\,1(R/1\,{\rm{pc}})^{1/2}\,\rm{km\,s^{-1}}$.
For slowly moving isolated BHs with velocity of $V\sim{c_{\rm{s}}}\sim10\,\rm{km\,s^{-1}}$, the contribution from the velocity fluctuation is evaluated as $\delta{V}/{V}\sim\delta{v_{\rm{turb}}}/c_{\rm{s}}\sim0.1\,(R/1\,{\rm{pc}})^{1/2}$.
At the Bondi radius, we see that the velocity fluctuation is much smaller than the density fluctuation.
Therefore, the disk radius is largely determined by the density fluctuation.

Then, at the centrifugal radius, the gas starts to circulate and forms an accretion disk.
The disk size is evaluated by the centrifugal radius as
\begin{eqnarray}
R_{\rm{d}}&:=&\frac{l^2}{GM}
   \label{disk radius}\\
&\simeq&4.7\times10^{9}\,\biggl(\frac{M}{10\,\Msun}\biggl)^{5/3}\biggl(\frac{V}{40\,\rm{km\,s^{-1}}}\biggl)^{-10/3}{\,\rm{cm}}.
   \label{disk radius2}
\end{eqnarray}
Since the mass accretion rate is much smaller than the Eddington rate, the accretion disk cools with not radiation but advection, so-called advection dominated accretion flow \citep[ADAF,][]{1994ApJ...428L..13N}.

When magnetic fields from the ISM thread the accretion disk, the magnetic braking effect may work and reduce the disk radius.
Since the efficiency of the braking effect sensitively depends on the geometry of the magnetic fields \citep[][and references therein]{1985A&A...142...41M}, it is difficult to study whether the magnetic braking affects the disk radius.
Here, we estimate the timescale of the angular momentum transfer by the Alfven crossing timescale at a given radius $R$, and compare it with the free fall timescale $t_{\rm{ff}}=\sqrt{R^3/GM}$.
At the Bondi radius $R_{\rm{B}}\simeq1.3\times10^{15}\,\rm{cm}$ for an isolated BHs with the velocity $V\simeq10\,\rm{km\,s^{-1}}$, each timescale is evaluated as $t_{\rm{A}}\simeq1.0\times10^{10}\,\rm{s}$ and $t_{\rm{ff}}\simeq1.3\times10^{9}\,\rm{s}$ for the magnetic field and the density at the Bondi radius of $B\simeq10\,\mu\rm{G}$ \citep{2010ApJ...725..466C} and $n\simeq10^2\,\rm{cm^{-3}}$.
These are the typical values in molecular clouds, where isolated BHs get enough accretion rate to power X-ray novae (see below).
Then, we can neglect the braking effect.
At the disk radius $R_{\rm{d}}\simeq4.8\times10^{11}\,\rm{cm}$, each timescale is estimated as $t_{\rm{A}}\simeq5.7\times10^{5}\,\rm{s}$ and $t_{\rm{ff}}\simeq9.2\times10^{3}\rm{\,s}$, where we use the flux freezing $BR^2=\rm{const}$ and the disk density given by Eq. \eqref{ss disk density}.
Therefore, it is unlikely that the braking effect transfers a significant amount of angular momentum, and reduces the disk radius.
In order to study this effect in detail, we may need to conduct a magnetohydrodynamic simulation, but it is the beyond of the scope of this work.

For a stationary accretion disk to exist, the infall time at the Bondi radius $t_{\rm{ff}}(R_{\rm{B}})$ should be shorter than the dynamical time of a BH to cross the Bondi radius $t_{\rm{dyn}}:=2R_{\rm{B}}/v$.
This condition is achieved as $t_{\rm{ff}}(R_{\rm{B}})/t_{\rm{dyn}}=v/2V\simeq0.35\lesssim1$, where we set the BH velocity is equal to the sound velocity $v\simeq{c_{\rm{s}}}$.

We calculate the mass accretion distribution function of Galactic isolated BHs.
The accreting ISM gas has several phases with different densities.
BHs also have velocity and mass distributions.
Therefore, we have to take into account for these statistical properties.
By using the normalized mass and accretion rate defined as $m:=M/\Msun$ and $\dot{m}:=\dot{M}/\dot{M}_{\rm{Edd}}$, the mass accretion distribution is given by \citep{2002MNRAS.334..553A,2017MNRAS.470.3332I}
\begin{eqnarray}
\frac{dN}{d\dot{m}}&=&N\int{dm}\frac{dp(m)}{dm}\int{dv}\frac{df(v)}{dv}\nonumber\\
&&\int{dn}\frac{d\xi(n)}{dn}h(m,v)\delta\biggl(\dot{m}(n,m,v)-\dot{m}\biggl),
   \label{LF}
\end{eqnarray}
where $N$, $dp(m)/dm$, $df(v)/dv$, $d\xi(n)/dn$ , and $h(m,v)$ are the total isolated BH number, the BH mass, the BH velocity and the ISM density distribution functions, and a correction factor due to the scale height of the ISM phase and the BH distribution.
We set the total number as $N=10^8$ \citep{1983bhwd.book.....S}.
We can integrate Eq. \eqref{LF} over $v$ and obtain
\begin{eqnarray}
\frac{dN}{d\dot{m}}=N\int{dm}\frac{dp(m)}{dm}\int{dn}\frac{df(v_0)}{dv}\frac{d\xi(n)}{dn}h(m,v_0)\frac{V^2|_{v=v_0}}{3v_0\dot{m}},
   \label{LF2}
\end{eqnarray}
where $v_0$ is given by 
\begin{eqnarray}
v_0^2=\biggl(\frac{GMc\eta\sigma_{\rm{T}}\mu{n}}{\dot{m}}\biggl)^{2/3}-c_{\rm{s}}^2.
\end{eqnarray}

\begin{table*}
\centering
\caption{The parameters for each ISM phase. For molecular clouds and cold \HI medium which have broad density distributions, the minimum $n_1$, maximum $n_2$ number densities, and power law index of the distributions $\beta$ are shown. For the other mediums, we show the typical density in the column 2. We represent the volume filling factors $\xi_0$, the sound velocities $c_{\rm{s}}$, and scale heights $H_{\rm{d}}$ in column 5, 6, 7, respectively. For the sound velocity, we include the contribution of the turbulent velocity.}
\label{table ism}
\begin{tabular}{llcclll}
\hline
phase & $n_1$ [$\rm{cm^{-3}}$] & $n_2$ [$\rm{cm^{-3}}$] & $\beta$ & $\xi_0$ & $c_{\rm{s}}$ [$\rm{km\,s^{-1}}$] & $H_{\rm{d}}$ [kpc]\\
\hline
Molecular cloud & $10^2$ & $10^5$ & 2.8 & $10^{-3}$ & 10 & 0.075\\
Cold \HI & $10$ & $10^{2}$ & 3.8 & 0.04 & 10 & 0.15\\
Warm \HI & $0.3$ & - & - & 0.35 & 10 & 0.5\\
Warm \HII & $0.15$ & - & - & 0.2 & 10 & 1\\
Hot \HII & $0.002$ & - & - & 0.4 & 150 & 3\\
\hline
\end{tabular}
\end{table*}

We briefly explain the BH mass, the BH velocity, and the ISM density distribution functions.
These functions are the same as those used in \cite{2017MNRAS.470.3332I}.
For the BH mass distribution function, we assume a Salpeter-like mass function as,
\begin{eqnarray}
\frac{dp(m)}{dm}=Cm^{-\gamma},\,\,\,(m_{\rm{min}}<m<m_{\rm{max}}),
\end{eqnarray}
where $\gamma=2.35$ and we normalize the mass function as $\int{dm}\frac{dp(m)}{dm}=1$ by setting $C=(\gamma-1)/(m_{\rm{min}}^{1-\gamma}-m_{\rm{max}}^{1-\gamma})$.
We set the upper and lower mass as $m_{\rm{min}}=5\,\Msun$ and $m_{\rm{max}}=15\,\Msun$.
The maximum BH mass is motivated by the stellar evolution calculation for the solar abundance \citep{2010ApJ...714.1217B}.
It should be noted that more massive BHs of $\sim50\,\Msun$ could form in low metallicity environments, as suggested by the GW observations \citep{2016ApJ...818L..22A}.
The number of the massive population could be comparable to that for the solar abundance because the duration of the low metal era is about tenth of the cosmic time, but the star formation rate is also ten times larger than now \citep{2014ApJ...781L..31S,2016A&A...589A..66H}.
Although considering their contribution is interesting, we do not take them into account because our result does not depend on the maximum BH mass so much.

In this work, we assume a Maxwellian velocity distribution as,
\begin{eqnarray}
\frac{df(v)}{dv}=\sqrt{\frac{2}{\pi}}\frac{v^2}{\sigma_{v}^3}\exp\biggl(-\frac{v^2}{2\sigma_{v}^2}\biggl),
\end{eqnarray}
where $\sigma_v$ is the velocity dispersion.
We set this value as $\sigma_{v}=40\,\rm{km/s}$.
This is motivated by the observed scale height of low mass X-ray binaries from the Galactic plane \citep{1996ApJ...473L..25W}.
Recently, \cite{2012MNRAS.425.2799R} have suggested that BHs also receives natal kicks at births and the kick velocity can be the same as those received by neutron stars ($\sim200-400\,\rm{km\,s^{-1}}$).
They also used the distance of the binaries from the Galactic plane to deduce the conclusion.
However, it should be noted that these arguments depend sensitively on the errors in the evaluated distances.
In order to get reliable distance or velocity values, we need more precise measurements such as astrometric observations \citep{2014PASA...31...16M}.
Such astrometric observations have been applied only for one BH X-ray binary system, Cyg X-1, which has a low proper velocity as $\sim20\,\rm{km\,s^{-1}}$ \citep{1998A&A...330..201C,2003Sci...300.1119M,2011ApJ...742...83R}.
 
We consider five types of the ISM phase: molecular clouds, cold \HI, warm \HI, warm \HII, and hot \HII mediums.
For molecular clouds and cold \HI medium, we adopt the power law distribution as \citep{1999ptep.proc...61B},
\begin{eqnarray}
\frac{d\xi(n)}{dn}=D\xi_0n^{-\beta},\,\,\,(n_1<n<n_2).
\end{eqnarray}
We normalize the function as $\int{dn}\frac{d\xi(n)}{dn}=\xi_0$ by choosing the constant $D=(\beta-1)/(n_{1}^{1-\beta}-n_2^{1-\beta})$, where $\xi_0$ is the volume filling factor of the phase in the Galactic volume.
The power law index $\beta$ and the upper and lower density $n_1$ and $n_2$ are shown in Table \ref{table ism}.
For the warm \HI, warm \HII, and hot \HII mediums, we do not consider the density distribution but assume a uniform density with their typical values.
Then, we use the delta function for these phases as
\begin{eqnarray}
\frac{d\xi(n)}{dn}=\xi_0\delta(n-n_1).
\end{eqnarray}
We take the value of the mean molecular weight as $\mu=2.82$ for the molecular clouds and $\mu=1.41$ for the other phases with the Milky Way abundance \citep{2008A&A...487..993K}.

\begin{figure}
\includegraphics[width=0.7\columnwidth,angle=270]{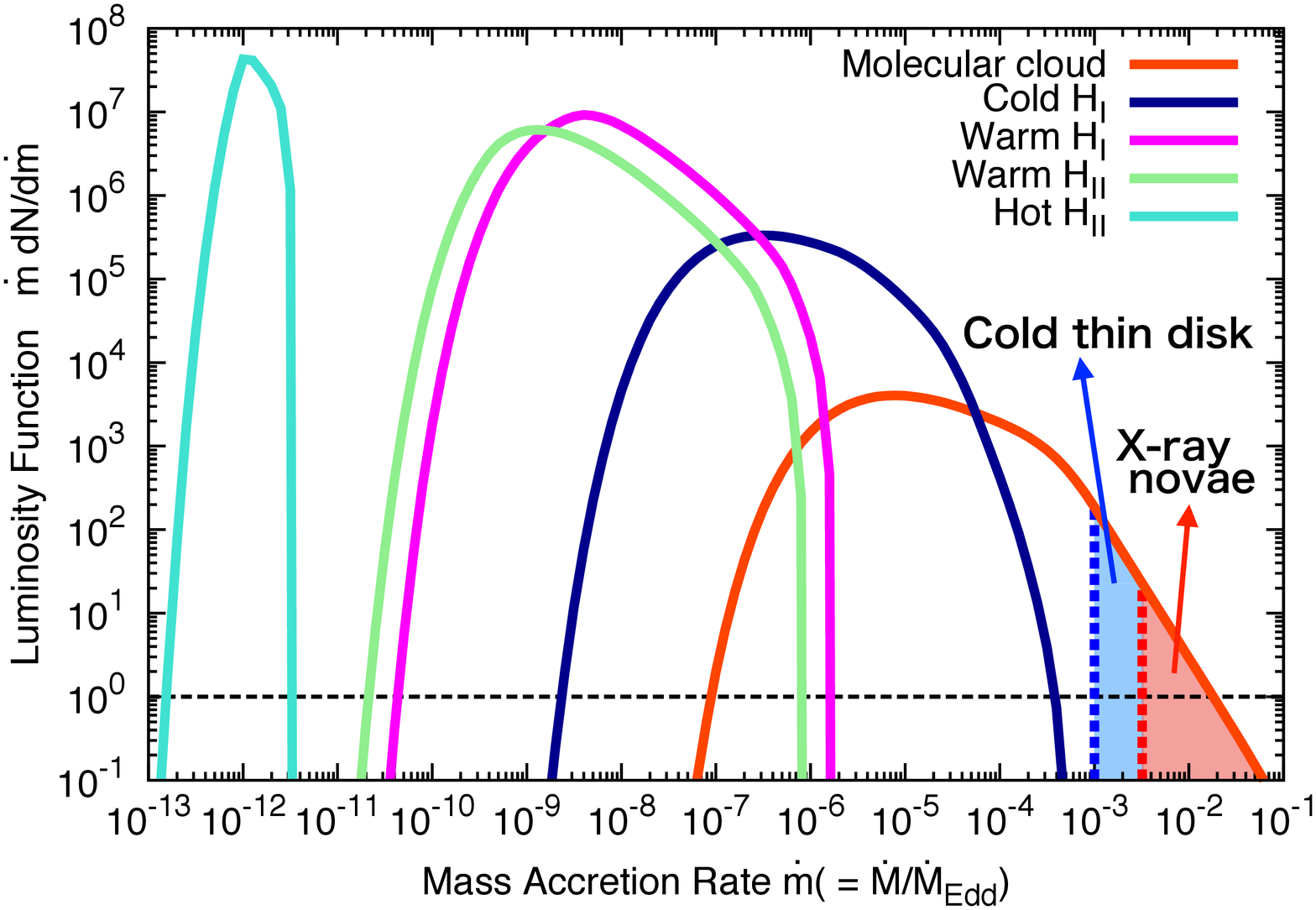}
\caption{The distribution of the normalized mass accretion rate $\dot{m}$. Each curve shows the accretion distribution in each medium.
The orange-red, dark-blue, magenta, light-green, and turquoise solid curves show the number distribution of isolated BHs in the molecular clouds, cold \HI, warm \HI, warm \HII, and hot \HII mediums, respectively.
The red and blue dashed lines represent the critical accretion rates required to produce X-ray novae (Eq. \ref{mdot_xrn}, for $m=15$) and to have a standard disk part (Eq. \ref{mdot_disk}, for $m=15$ and $n=10^{2}\,\rm{cm^{-3}}$), respectively.
}
\label{fig mf}
\end{figure}

We comment the uncertainty of the density distribution function of molecular clouds because molecular clouds are the main site where isolated BHs launch X-ray novae.
The power law distribution is obtained by the mass function of molecular clouds $dN_{\rm{cl}}/dM_{\rm{cl}}\propto{M_{\rm{cl}}}^{-p}$, where $N_{\rm{cl}}$ and $M_{\rm{cl}}$ are the number of molecular clouds and the molecular cloud mass, and the density and cloud size relation \cite[so-called Larson's third law,][]{1981MNRAS.194..809L}, $nR_{\rm{cl}}\propto{R_{\rm{cl}}}^{q}$, where $R_{\rm{cl}}$ is the molecular cloud size.
While the observed typical indices of $p=1.6$ \citep{1997ApJ...476..166W} and $q=0$ \citep{1987ApJ...319..730S} give $\beta=2.8$, the inaccuracy of these index value may also cause an uncertainty of our result. 
By the observations, the indices are determined within the range of $p\simeq1.5-1.8$ \citep{1998A&A...329..249K,2005PASP..117.1403R} and $q\simeq0-0.1$ \citep{2009ApJ...699.1092H}, which result in the index $\beta\simeq3.0-2.4$.
However, this change of $\beta$ increases or decreases only the maximum accretion rate ($\dot{m}\simeq2\times10^{-2}$, see Fig \ref{fig mf}) and dose not change seriously the total number of BHs which power X-ray novae.
Therefore, our main results such as the event rate does not sensitively depend on the uncertainty of the density distribution functions.
In addition to the density distribution, the turbulent velocity distribution may change the accretion distribution.
Note that we include this contribution into the sound velocity (see the caption in Table \ref{table ism}).  
The Larson's first law shows that the smaller cloud has the smaller turbulent velocity, which makes the BH accretion rate large.
However, according to the Larson's third law, these clouds have also large density and only change the maximum accretion rate.

Finally, we explain the correction factor $h(m,v)$.
For a given BH velocity, we can evaluate the BH's scale height by assuming the Galactic potential.
We use the following simple potential model as,
\begin{eqnarray}
v_z^2&=&v^2/3\\
\frac{1}{2}v_z^2&=&\Phi[H(v_z)]\\
\frac{\Phi(z)}{2\pi{G}}&=&K\biggl(\sqrt{z^2+Z^2}-Z\biggl)+Fz^2,
   \label{galactic potential}
\end{eqnarray}
where $Z=180\,\rm{pc}$, $K=48\,\Msun\,\rm{pc^{-2}}$, and $F=0.01\,\Msun\,\rm{pc^{-3}}$ \citep{1989MNRAS.239..571K,1989MNRAS.239..605K}.
When the derived scale height $H(v_z)$ is larger than the scale height of the ISM phase $H_{\rm{d}}$, we correct the count by multiplying $H_{\rm{d}}/H(v_z)$.
Then, the correction factor is given by
\begin{eqnarray}
h(m,v)={\rm{min}}\biggl[1,\frac{H_{\rm{d}}}{H(v_z)}\biggl].
\end{eqnarray}

We compile the distribution functions discussed above, and integrate Eq. \eqref{LF2} numerically.
In Fig \ref{fig mf}, we show the mass accretion distribution of Galactic isolated BHs.
The vertical and horizontal axises show the number of BHs and the normalized mass accretion rate, respectively.
The normalized accretion rate relates with the BH mass, the BH velocity, and the ISM density as
\begin{eqnarray}
\dot{m}\simeq5.8\times10^{-8}\,\biggl(\frac{M}{10\,\Msun}\biggl)\biggl(\frac{V}{40\,{\rm{km\,s^{-1}}}}\biggl)^{-3}\biggl(\frac{n}{1\,{\rm{cm^{-3}}}}\biggl),
   \label{mdot estimate}
\end{eqnarray}
where we use Eqs. \eqref{bondi rate} and \eqref{eddington rate}.
The orange-red, dark-blue, magenta, light-green, and turquoise curves show the number distributions of isolated BHs in the molecular clouds, cold \HI, warm \HI, warm \HII, and hot \HII mediums, respectively.

Let us discuss the shape of the distribution functions in Fig \ref{fig mf}.
The peak value of the distribution in each ISM medium is roughly evaluated by $N_{\rm{peak}}\sim{N}\xi_0h$.
For the hot \HII medium, the distribution shows a peaky shape.
This is because the hot \HII medium has a larger sound velocity $c_{\rm{s}}$ than the typical BH velocity $v\sim{\sigma_{\rm{v}}}$, and the distribution reflects only the BH mass distribution.
According to the Galactic potential \eqref{galactic potential}, the BH scale height with the velocity $v\sim\sigma_{\rm{v}}=40\,\rm{km\,s^{-1}}$ is estimated as $H(v_{\rm{z}})\simeq0.3\,\rm{kpc}$.
On the other hand, the scale height of the hot \HII medium is larger than the BH scale height (see Table \ref{table ism}), and the correction factor reduces to unity $h=1$.
Then, the peak value of the distribution of the hot \HII medium is given as $N_{\rm{peak}}\sim4\times10^7\,(N/10^8)(\xi_0/0.4)(h/1)$.

For the warm \HI and \HII mediums, the mass accretion distributions have a broader shape than that of the hot \HII medium because the sound velocities of these mediums are as large as that of the BH velocity, $c_{\rm{s}}\lesssim{v}\sim\sigma_{\rm{v}}$.
Therefore, the broadness reflects the velocity distribution.
These mediums also have larger scale heights than the BH scale height, and have no correction ($h\sim1$).
However, due to the large dispersion, the peak values of the distributions are a bit smaller than the values estimated by $\sim{N}\xi_0$. 

The accretion distributions of the molecular clouds and the cold \HI medium show tails extending to the large accretion rate ($\dot{m}\gtrsim10^{-5}$ and $10^{-3}$, for the cold \HI medium and the molecular clouds, respectively).
These tails are attributed to the density distribution.
To estimate the peak values of the distributions, we should take the correction factor $h$ into account, because the molecular clouds and the cold \HI medium have smaller scale heights than that of BHs, $H_{\rm{d}}<H(v_{\rm{z}})$, which requires corrections of $h\sim0.1-0.3$.

\section{X-ray Novae produced by Isolated BHs}\label{X-ray Novae produced by Isolated BHs}
We discuss the possibility that isolated BHs produce X-ray transient events, such as X-ray novae.
In section \ref{Accretion Rate Distribution of isolated BHs}, we consider that the accretion disks formed around isolated BHs are ADAFs.
However, when the accretion rate is larger than a critical value, the radiative cooling overcomes the advective one and the ADAF becomes a standard disk \citep{1995ApJ...438L..37A}.
Furthermore, if the standard disk has a low temperature region enough to allow hydrogen to recombine, the disk suffers from the hydrogen-ionization instability and produces an X-ray transient event.

Before discussing the accretion disk structure in detail, we check whether our idea passes the observational constraints.
As shown in Fig. \ref{fig mf} and we discuss below, X-ray novae from isolated BHs occur only in molecular clouds because a large accretion rate is necessary.
Then, if an X-ray nova is actually powered by an isolated BH, we should detect it in the same direction of a molecular cloud.\footnote{Since it is difficult to measure the distance to an X-ray nova only by X-ray observations, we cannot study whether the X-ray nova actually reside in a molecular cloud. Therefore, we only discuss the coincidence of the directions of X-ray novae and molecular clouds.}
We check an X-ray nova catalog \citep{2016A&A...587A..61C} and find 17 events in the Galactic plane ($b\lesssim1.5\,\rm{deg}$) among 59 X-ray novae ever detected.
Comparing the CO emission map in the Galactic plane,\footnote{\url{https://www.cfa.harvard.edu/rtdc/CO/IndividualSurveys/}} we find CO emission signatures in the same directions of 16 X-ray novae.
Furthermore, the 16 X-ray novae show large column densities of $\sim10^{22}\,\rm{cm^{-2}}$ in the soft X-ray spectra, which also support that the 16 X-ray novae are actually accompanied with molecular clouds (see also section \ref{Discussion}).
We list the 16 candidate X-ray novae in Table \ref{table candidate}.
The column density and direction are taken from the catalog of \cite{2016A&A...587A..61C} and references therein.
Therefore, we conclude $16/59\sim0.27$ of the observed X-ray novae are potentially produced by isolated BHs in molecular clouds.

\begin{table*}
\centering
\caption{Candidates X-ray novae powered by isolated BHs in molecular clouds. ID number corresponds to that used in \citep{2016A&A...587A..61C}. The column density and direction are taken from \citep{2016A&A...587A..61C} and references therein. The column density is evaluated by the absorption signature in the soft X-ray spectra, except for ID47, 28, 22, 11, and, 2, for which the optical extinction formula $N_{\rm{H}}=5.8\times10^{21}E(B-V)\,\rm{cm^{-2}}$ is used \citep{2011piim.book.....D}. In the same direction of each event, we find that there is a molecular cloud by checking the CO emission map. NIR counterparts are detected only for ID59, 35, and 34.}
\label{table candidate}
\begin{tabular}{llrccc}
\hline
ID & name &$N_{\rm{H}}$ [$\rm{cm^{-2}}$]& direction $(l,b)$& NIR source \\
\hline
59 & IGR J17454-2919 & $1.0$-$1.2\times10^{23}$ & $359.6444,-00.1765$ &$\checkmark$ \\
56 & SWIFT J1753.7-2544 & $5.8\times10^{22}$ & $003.6476,+00.1036$ &\\ 
51 & MAXI J1543-564 &$1.4\times10^{22}$& $325.0855,-01.1214$ &\\
47 & XTE J1652-453 &$5.1\times10^{22}$& $340.5297,-00.7867$ &\\
44 & SWIFT J174540.2-290005 &$7.6\times10^{22}$& $359.9495,-00.0431$&\\
43 & IGR J17497-2821 &$4.8\times10^{22}$& $000.9531,-00.4527$\\
35 & XTE J1908+094 & $2.3\times10^{22}$ & $043.2615,+00.4377$ &$\checkmark$\\
34 & SAX J1711.6-3808 & $2.8\times10^{22}$ & $348.4200,+00.7880$ &$\checkmark$\\
28 & XTE J1748-288 &$5.8\times10^{22}$& $000.6756,-00.2220$ &\\
25 & GRS 1737-31 &$6.0\times10^{22}$& $357.5880,-00.0990$ &\\
24 & GRS 1739-278 &$1.6\times10^{22}$& $000.6721,+01.1758$ &\\
23 & XTE J1856+053 &$4.5\times10^{22}$& $038.2690,+01.2720$ &\\
22 & GRS 1730-312 &$3.2\times10^{22}$& $356.6877,+01.0065$ &\\
11 & EXO 1846-031 &$3.7\times10^{22}$& $029.9585,-00.9177$ &\\
8 & H 1743-322 & $2.2\times10^{22}$ & $357.2552,-01.8330$ &\\
2 & 4U 1630-472 &$2.4\times10^{22}$& $336.9112,+00.2503$ &\\
\hline
\end{tabular}
\end{table*}

When the mass accretion rate is large, an ADAF can accompany a standard disk at the outer part of the disk.
This is because at a large radius from the central BH, the radiative cooling rate gets comparable to the advective cooling and viscosity heating rates.
We can see this easily by considering the energy equation of ADAFs as 
\begin{eqnarray}
(1-f)q_{\rm{vis}}\simeq{q_{\rm{bre}}}
   \label{energy eq}
\end{eqnarray}
where $q_{\rm{vis}}$ and $q_{\rm{bre}}$ are the viscous  heating and the bremsstrahlung cooling rates per volume, respectively.
We also assume that the advective cooling rate is approximately given by the viscous heating rate as $q_{\rm{adv}}\simeq{f}q_{\rm{vis}}$, where $f$ is the efficiency parameter and we set $f=0.5$ for a fiducial value at the transition region to a standard disk \citep{1995ApJ...452..710N}.
Then, the left hand side of Eq. \eqref{energy eq} represents a net heating rate.
By using a self-similar solution of ADAF, we can show that the heating rate depends on radius $r$ as $q_{\rm{vis}}\propto{r^{-4}}$ \citep{1994ApJ...428L..13N,1995ApJ...452..710N}, where we normalize a radius by the Schwarzschild radius $R_{\rm{S}}=2GM/c^2$ as $r:=R/R_{\rm{S}}$.
It should be also noted again that we normalize the mass and accretion rate by the solar mass and the Eddington rate as $m:=M/\Msun$ and $\dot{m}:=\dot{M}/\dot{M}_{\rm{Edd}}$.
The bremsstrahlung cooling rate is given by \cite{1982ApJ...258..335S}, and we can show the radius dependence of the cooling rate as $q_{\rm{bre}}\propto{r^{-7/2}}$, where we use the ADAF self-similar solution and assume that electrons have a non-relativistic virial temperature of ions\footnote{At the outer part of the ADAF ($r\gtrsim100$), the Coulomb collision works well and makes the electron temperature equal to the ion temperature.} as \citep{1995ApJ...452..710N,Matsumoto}
\begin{eqnarray}
T_{\rm{e}}\simeq3.2\times10^{12}\,\beta_{\rm{ADAF}}\,c_3\,r^{-1}\rm{\,K}.
\end{eqnarray}
The constants $\beta_{\rm{ADAF}}$ and $c_3$ are the parameters of the ADAF self-similar solution (see below Eq. \eqref{critical mdot} for more detailed explanation). 
Then, at a large radius, the radiative cooling rate becomes comparable to the viscous heating rate, and the ADAF becomes a standard disk.
Equation \eqref{energy eq} is rewritten to the condition for the critical accretion rate at a given radius by using the self-similar solution as \citep{1995ApJ...438L..37A,1995ApJ...452..710N},
\begin{eqnarray}
\dot{m}_{\rm{crit}}&\simeq&2.3\,\biggl(\frac{m_{\rm{u}}}{\ME}\biggl)^{1/2}\frac{\eta}{\alpha_{\rm{f}}}\,\epsilon^\prime(1-f)c_1^2c_3\alpha^2\beta_{\rm{ADAF}}^{-1/2}r^{-1/2}\\
&\simeq&1.4\times10^{3}\,\epsilon^\prime(1-f)c_1^2c_3\alpha^2\beta_{\rm{ADAF}}^{-1/2}r^{-1/2}
   \label{critical mdot}\\
&\simeq&4.4\times10^{-1}\biggl(\frac{\epsilon^\prime}{0.63}\biggl)\biggl(\frac{1-f}{0.5}\biggl)\nonumber\\
&&\biggl(\frac{c_1}{0.48}\biggl)^2\biggl(\frac{c_3}{0.32}\biggl)\biggl(\frac{\alpha}{0.1}\biggl)^2\biggl(\frac{\beta_{\rm{ADAF}}}{0.5}\biggl)^{-1/2}r^{-1/2},
\end{eqnarray}
where $\ME$, $\alpha_{\rm{f}}$, $\alpha$ and $\beta_{\rm{ADAF}}$ are the electron mass, the fine structure constant, the viscosity parameter \citep{1973A&A....24..337S} and the ratio of the gas pressure to the total pressure, respectively.
For fiducial values, we set $\alpha=0.1$ and $\beta_{\rm{ADAF}}=0.5$.
The dimensionless constants $\epsilon^{\prime}$, $c_1$, and  $c_3$ appear in the derivation of the ADAF self-similar solution as functions of $f$, $\alpha$, and $\beta_{\rm{ADAF}}$ \citep{1994ApJ...428L..13N,1995ApJ...452..710N}.
The notation of these constants is the same as given in \cite{1994ApJ...428L..13N,1995ApJ...452..710N}, except for $\beta_{\rm{ADAF}}$ (where $\beta_{\rm{ADAF}}$ is written as $\beta$).

If $\dot{m}\gtrsim\dot{m}_{\rm{crit}}$, the radiative cooling overcomes the advective cooling and no longer the ADAF part exists.
Observations of BH X-ray binaries also suggest an outer standard accretion disk truncated in the inner part by radiatively inefficient accretion flow, which can explain some properties of BH X-ray binaries such as the spectral transition between the high-soft state and the low-hard state \citep{1997ApJ...489..865E,2001NewAR..45..449L,2007A&ARv..15....1D}.

We estimate how many isolated BHs have the standard disk part.
We define the transition radius $r_{\rm{tr}}(=R_{\rm{tr}}/R_{\rm{S}})$ by $\dot{m}=\dot{m}_{\rm{crit}}(r=r_{\rm{tr}})$, where the ADAF part begins to cool radiatively and makes a transition to the standard disk outer-part.
By using Eqs. \eqref{mdot estimate} and \eqref{critical mdot}, we obtain the transition radius as,
\begin{eqnarray}
r_{\rm{tr}}\simeq1.5\times10^{9}\,\biggl(\frac{m}{10}\biggl)^{-2}\biggl(\frac{V}{40\,{\rm{km\,s^{-1}}}}\biggl)^6\biggl(\frac{n}{10^2{\rm{\,cm^{-3}}}}\biggl)^{-2}.
   \label{r_tr}
\end{eqnarray}
It should be noted that when we discuss the BHs in molecular clouds, we should use the mean molecular weight of $\mu=2.82$ in Eq. \eqref{mdot estimate}.
We use the fiducial values for the ADAF parameters in Eq. \eqref{critical mdot}.
In the following, we also use the fiducial values and do not show the dependences of ADAF parameters explicitly.  
For a disk to have a standard disk part, the transition radius should be smaller than the disk radius $r_{\rm{d}}:=R_{\rm{d}}/R_{\rm{S}}$.
With Eq. \eqref{disk radius}, this condition is rewritten to a condition for the velocity $V$ as,
\begin{eqnarray}
V\lesssim9.2\,\biggl(\frac{m}{10}\biggl)^{2/7}\biggl(\frac{n}{10^2{\rm{\,cm^{-3}}}}\biggl)^{3/14}\,{\rm{km\,s^{-1}}}.
   \label{v1}
\end{eqnarray}
Since the velocity is $V>c_{\rm{s}}=10\,\rm{km\,s^{-1}}$, to satisfy the above condition, the BH mass and the density should be larger than the minimum values, $m>5$ and $n>10^2\,\rm{cm^{-3}}$.
Hereafter, we use $m=15$ for the fiducial value.
By substituting Eq. \eqref{v1} into Eq. \eqref{mdot estimate}, we obtain a required minimum accretion rate for an isolated BH to have a standard disk part as
\begin{eqnarray}
\dot{m}\gtrsim\dot{m}_{\rm{disk}}&=&1.0\times10^{-3}\,\biggl(\frac{m}{15}\biggl)^{1/7}\biggl(\frac{n}{10^2{\rm{\,cm^{-3}}}}\biggl)^{5/14}\nonumber\\
&&\biggl(\frac{\epsilon^\prime}{0.63}\biggl)^{9/14}\biggl(\frac{1-f}{0.5}\biggl)^{9/14}\biggl(\frac{c_1}{0.48}\biggl)^{9/7}\nonumber\\
&&\biggl(\frac{c_3}{0.32}\biggl)^{9/14}\biggl(\frac{\alpha}{0.1}\biggl)^{9/7}\biggl(\frac{\beta_{\rm{ADAF}}}{0.5}\biggl)^{-9/28},
   \label{mdot_disk}
\end{eqnarray}
where we restore the ADAF parameter dependences.
In Fig \ref{fig mf}, we show the minimum accretion rate with a blue dashed line for $m=15$ and $n=10^2\,\rm{cm^{-3}}$.

Next, we discuss the possibility that hydrogen recombines in the thin disk part.
We firstly compare the recombination temperature with the temperature in the disk for simplicity, and later discuss the details.
The Saha equation gives the hydrogen recombination temperature in the thin disk.
When hydrogen is partially ionized, their ionization state determines the abundance of $\rm{H^-}$ ion (i.e., opacity) (The metal abundance may not change results so much as long as hydrogen mainly supplies electrons).
The Saha equation gives the ratio of the ionized hydrogen number density $n_{\rm{H^+}}$ to the neutral hydrogen density $n_{\rm{H}}$ as
\begin{eqnarray}
\frac{n_{\rm{H^+}}n_{\rm{e}}}{n_{\rm{H}}}=\frac{(2\pi\ME\KB{T})^{3/2}}{h^3}\exp\biggl({-\frac{\chi_{\rm{H}}}{\KB{T}}}\biggl),
\label{saha}
\end{eqnarray}
where $n_{\rm{e}}$, $\KB$, $h$, and $\chi_{\rm{H}}=13.6\,\rm{eV}$ are the electron number density, the Boltzmann constant, the Planck constant, and the hydrogen ionization energy.
With the definition of the degree of ionization $x:=n_{\rm{e}}/(n_{\rm{H}}+n_{\rm{H^+}})$, and the charge neutrality $n_{\rm{e}}=n_{\rm{H^+}}$, Eq. \eqref{saha} is rewritten as 
\begin{eqnarray}
\frac{x^2}{1-x}=\frac{(2\pi\ME\KB{T})^{3/2}}{n_{\rm{H,tot}}h^3}\exp\biggl({-\frac{\chi_{\rm{H}}}{\KB{T}}}\biggl),
   \label{saha2}
\end{eqnarray}
where $n_{\rm{H,tot}}=n_{\rm{H}}+n_{\rm{H^+}}$ is the total number density of hydrogen.
We simply define the partially ionized state of the hydrogen as $x=0.5$, which results in the value of the right hand side of Eq. \eqref{saha2} of $0.5$.
The temperature and the mass density are given by formulae of the standard disk with gas pressure and free-free absorption \citep{1973A&A....24..337S,2008bhad.book.....K} as 
\begin{eqnarray}
T&=&1.4\times10^{8}\,\alpha^{-1/5}\,m^{-1/5}\,\dot{m}^{3/10}\,r^{-3/4}\,\rm{K}
   \label{ss disk temperature}\\
\rho&=&1.7\times10^{2}\,\alpha^{-7/10}\,m^{-7/10}\,\dot{m}^{11/20}\,r^{-15/8}\,\rm{g\,cm^{-3}}.
   \label{ss disk density}
\end{eqnarray}
The total number density of hydrogen is obtained by $n_{\rm{H,tot}}=\rho{X}/\MP$, where $X=0.7$ is the mass fraction of hydrogen for the solar abundance.
Substitution of above expressions for Eq. \eqref{saha2} yields the radius where hydrogen begins to recombine as
\begin{eqnarray}
r&\simeq&8.3\times10^3\biggl(7.4+\ln\biggl[\alpha^{2/5}m^{2/5}\dot{m}^{-1/10}r^{3/4}\biggl]\biggl)^{4/3}\nonumber\\
&&\alpha^{-4/15}m^{-4/15}\dot{m}^{2/5}\\
&\simeq&1.3\times10^{4}\,\biggl(\frac{\alpha}{0.1}\biggl)^{-4/15}\biggl(\frac{m}{15}\biggl)^{-4/15}\biggl(\frac{\dot{m}}{10^{-3}}\biggl)^{2/5}.
   \label{recombination radius}
\end{eqnarray}
From the first to the second line, we substitute the arguments in the natural logarithm for the typical parameter values $\alpha=0.1$,\footnote{In the hydrogen-ionization instability, the viscous parameter $\alpha$ is assumed to change between the hot and cold branches of the S-curve, in order to explain the outburst amplitude \citep{1983PASJ...35..377M,1984A&A...132..143M,1984AcA....34..161S}.
The change of the viscosity parameter is also supported by the detailed analysis of the dwarf-nova observations \citep{2012A&A...545A.115K}, and magnetohydrodynamic simulations of accretion disks \citep{2016MNRAS.462.3710C}. In our estimate, we use the viscous parameter $\alpha=0.1$ in the hot branch.} $m=15$, $\dot{m}=10^{-3}$, and $r=10^4$ suggested by Eq. \eqref{mdot_disk}.
The necessary accretion rate for hydrogen recombination is also obtained by solving Eq. \eqref{recombination radius} for the mass accretion rate as \footnote{When hydrogen recombines and makes the opacity large, convection develops in accretion disks, which decreases the vertical temperature gradient \citep{1982A&A...106...34M,1984ApJS...55..367C}. Therefore, if the convective motion reaches the midplane of the disk, the standard disk formula \eqref{ss disk temperature} overestimates the disk temperature. However, \cite{1984ApJS...55..367C} shows that the convection does not reach the midplane at the maximum accretion rate of the S-curve middle branch.}
\begin{eqnarray}
\dot{m}\simeq8.0\times10^{-15}\biggl(\frac{\alpha}{0.1}\biggl)^{2/3}r^{5/2}m^{2/3}.
   \label{analytical maximum rate}
\end{eqnarray}
By substituting the above radius into the temperature \eqref{ss disk temperature}, we obtain 
\begin{eqnarray}
T\simeq1.3\times10^4\,\rm{K}.
   \label{H recombination temperature}
\end{eqnarray}
This disk temperature corresponds to the effective temperature of 
\begin{align}
T_{\rm{eff}}=\biggl[\frac{3GM\dot{M}}{8\pi\sigma_{\rm{SB}}R^3}\biggl]^{1/4}\simeq4.5\times10^3\,\biggl(\frac{\alpha}{0.1}\biggl)^{1/5}\biggl(\frac{m}{15}\biggl)^{-1/20}\biggl(\frac{\dot{m}}{10^{-3}}\biggl)^{-1/20}\,\rm{K},
\end{align}
where $\sigma_{\rm{SB}}$ is the Stefan-Boltzmann constant. 
The hydrogen recombination temperature \eqref{H recombination temperature} should be compared with the temperature of the disks which we consider.

We roughly estimate the temperature of the thin disk part and show that the disk has a low temperature region where hydrogen can recombine.
We use the standard disk formulae in eq. \eqref{ss disk temperature}, and evaluate the disk temperature as 
\begin{eqnarray}
T\simeq2.8\times10^3\,\biggl(\frac{m}{15}\biggl)^{-1/5}\biggl(\frac{\dot{m}}{10^{-3}}\biggl)^{10/3}\biggl(\frac{r_{\rm{d}}}{10^5}\biggl)^{-3/4}\biggl(\frac{\alpha}{0.1}\biggl)^{-1/5}\,{\rm{K}},
   \label{disk temperature}
\end{eqnarray}
where we use the accretion rate and the disk radius of $\dot{m}=10^{-3}$ and $r_{\rm{d}}=10^5$ which corresponds to BHs with $m=15$ and $V=10\,\rm{km\,s^{-1}}$.
This value is lower than the hydrogen recombination temperature in Eq. \eqref{H recombination temperature}.
Therefore, the thin disk part of isolated BHs with accretion rate $\dot{m}\gtrsim10^{-3}$ has the region where hydrogen begins to recombine and makes the disk unstable.

We study the condition of hydrogen recombination more precisely than the above discussion, in particular, on the basis of the disk instability theory \citep{1996ApJ...464L.139V}.
For an instability to occur, the mass accretion rate should be in the middle branch of the S-curve at some radius.
The maximum and minimum accretion rates of the branch are calculated numerically, and fitting formulae are given by \cite{2008A&A...486..523L} as\footnote{It should be noted that \cite{2008A&A...486..523L} mainly studies the stability of helium accretion disks. However, in the Appendix of their paper, they show the fitting formulae for the solar abundance disk, which we use in this work.} 
\begin{eqnarray}
\dot{m}_{\rm{crit}}^{+}&=&6.36\times10^{-15}\,\biggl(\frac{\alpha}{0.1}\biggl)^{-0.01}r^{2.64}m^{0.75},
   \label{maximum rate}\\
\dot{m}_{\rm{crit}}^{-}&=&3.91\times10^{-15}\,\biggl(\frac{\alpha}{0.1}\biggl)^{0.01}r^{2.58}m^{0.73},
   \label{minimum rate}
\end{eqnarray}
where we rewrite the radius and mass from the original notations in \cite{2008A&A...486..523L} to the normalized ones, and keep the viscosity parameter $\alpha$ explicitly.
It should be noted the numerically obtained maximum accretion rate \eqref{maximum rate} is roughly reproduced by our analytical formula \eqref{analytical maximum rate} except for the dependence on the viscosity parameter $\alpha$.
Since the critical rates are increasing functions of radius, for an unstable annulus to exist, the mass accretion rate should be smaller than the maximum rate \eqref{maximum rate} at the disk radius, $\dot{m}<\dot{m}_{\rm{crit}}^{+}(r=r_{\rm{d}})$, and should also be larger than the minimum rate \eqref{minimum rate} at the transition radius, $\dot{m}>\dot{m}_{\rm{crit}}^{-}(r=r_{\rm{tr}})$.
For the first condition, the maximum rate at the disk radius is given by
\begin{eqnarray}
\dot{m}_{\rm{crit}}^{+}(r=r_{\rm{d}})&=&5.7\,\biggl(\frac{\alpha}{0.1}\biggl)^{-0.01}\biggl(\frac{m}{15}\biggl)^{2.51}\biggl(\frac{V}{10\,\rm{km\,s^{-1}}}\biggl)^{-8.8},
\end{eqnarray}
where we use Eq. \eqref{disk radius} for the disk radius.
We see that the first condition is easily satisfied.
For the second condition, we substitute Eq. \eqref{r_tr} to the minimum rate and obtain
\begin{eqnarray}
\dot{m}_{\rm{crit}}^{-}(r=r_{\rm{tr}})&=&7.3\times10^{-1}\,\biggl(\frac{\alpha}{0.1}\biggl)^{0.01}\biggl(\frac{m}{15}\biggl)^{-4.43}\nonumber\\
&&\biggl(\frac{V}{10\,{\rm{km\,s^{-1}}}}\biggl)^{15.48}\biggl(\frac{n}{10^2\,{\rm{cm^{-3}}}}\biggl)^{-5.16}.
\end{eqnarray}
At first glance, this condition does not seem to be satisfied.
However, for BHs with $\dot{m}\gtrsim10^{-3}$, the density is larger than $n=10^2\,{\rm{cm^{-3}}}$, which decrease the minimum rate from the above value.
More precisely, to satisfy the second condition, an inequality $(m/15)^{0.88}(n/10^2{\,\rm{cm^{-3}}})(V/10{\,\rm{km\,s^{-1}}})^{-3}>2.9$ should hold.
Then, by noting that $\dot{m}\propto{mV^{-3}n}$, we obtain the condition for isolated BHs to satisfy the second condition
\begin{eqnarray}
\dot{m}\gtrsim\dot{m}_{\rm{XRN}}&\simeq&3.2\times10^{-3}\,\biggl(\frac{m}{15}\biggl)^{0.12}\biggl(\frac{\epsilon^\prime}{0.63}\biggl)^{0.84}\biggl(\frac{1-f}{0.5}\biggl)^{0.84}\nonumber\\
&&\biggl(\frac{c_1}{0.48}\biggl)^{1.68}\biggl(\frac{c_3}{0.32}\biggl)^{0.84}\biggl(\frac{\alpha}{0.1}\biggl)^{1.68}\biggl(\frac{\beta_{\rm{ADAF}}}{0.5}\biggl)^{-0.42},
   \label{mdot_xrn}
\end{eqnarray}
where we restore the ADAF parameter dependences. 
In Fig. \ref{fig mf}, we also show the required accretion rate for $m=15$ with a red dashed line.

We conclude that isolated BHs with a mass accretion rate larger than the critical rates \eqref{mdot_disk} and \eqref{mdot_xrn} suffer from the hydrogen-ionization instability and cause transient events such as X-ray novae. 
In Fig \ref{fig mf}, we show the population which causes transients with a red shaded region.
We see that the number of the isolated BHs which can produce X-ray novae-like transients by the ionization instability is about $N_{\rm{BH}}\sim20$.
BHs in a blue shaded region have mass accretion rates of $\dot{m}_{\rm{disk}}<\dot{m}<\dot{m}_{\rm{crit}}^{-}(r=r_{\rm{tr}})$.
In this case, the whole region of the thin disk parts is always in the cold branch and stable \citep{1999ApJ...513..811M}.

The X-ray novae powered by isolated BHs show similar timescale and luminosity to the observed ones.
We briefly estimate the rise and decay timescales and the luminosity of the X-ray novae.
Before the detailed estimation, we remark that the accretion rate ($\dot{m}\gtrsim3\times10^{-3}$, see also Fig. \ref{fig mf}) and the disk radius ($R_{\rm{d}}\sim10^{11}\,\rm{cm}$ for $V\sim10\,\rm{km\,s^{-1}}$) of the isolated BHs are comparable to those of the X-ray binary system \citep{vanParadijs1996,Coriat+2012,2006ARA&A..44...49R}, which means that the accretion disks in both systems have common properties and show the common outburst signatures.
The rise-time is evaluated by the timescale for which a heating-front with velocity $\sim\alpha{c_{\rm{s,disk}}}$ propagates the disk radius \citep{2001NewAR..45..449L}, as $t_{\rm{rise}}\sim{}R_{\rm{d}}/\alpha{c_{\rm{s,disk}}}\sim10\,{\rm{days}}\,(R_{\rm{d}}/10^{11}{\rm{\,cm}})(\alpha/0.1)^{-1}(c_{\rm{s,disk}}/10\,\rm{km\,s^{-1}})^{-1}$, where $c_{\rm{s,disk}}$ is the sound velocity of the disk.
The decay-time is estimated by the viscous timescale as $t_{\rm{decay}}\sim{t_{\rm{vis}}}\sim{R_{\rm{d}}^2}/3\nu_{\rm{vis}}$, where $\nu_{\rm{vis}}\simeq\alpha{c_{\rm{s,disk}}}H_{\rm{disk}}$ is the kinetic viscosity, and $H_{\rm{disk}}$ is the disk scale height at the disk radius \citep{1998MNRAS.293L..42K}.
Then, the decay timescale is given by $t_{\rm{decay}}\sim40\,{\rm{days}}\,(R_{\rm{d}}/10^{11}{\,\rm{cm}})^2(\alpha/0.1)^{-1}(c_{\rm{s,disk}}/10\,{\rm{km\,s^{-1}}})^{-1}(H_{\rm{disk}}/0.1R_{\rm{d}})^{-1}$.
Finally, since the outburst luminosity is evaluated by $L_{\rm{burst}}\sim\eta\dot{M}_{\rm{burst}}c^2$.
The accretion rate is given by $\dot{M}_{\rm{burst}}\sim{M_{\rm{d}}}/t_{\rm{vis}}$, where $M_{\rm{d}}\sim\pi{\Sigma}R_{\rm{d}}^2$ is the disk mass at the beginning of the outburst and $\Sigma$ is the surface density at the cold branch of the S-curve.
The surface density is obtained by solving the disk structure, but we evaluate it by using the fitting formula developed by \cite{2008A&A...486..523L} (their Eq. A.1) as $\Sigma\sim6\times10^2\,{\rm{g\,cm^{-2}}}\,(\alpha/0.1)^{-0.83}(R_{\rm{d}}/10^{11}\,{\rm{cm}})^{1.18}(M_{\rm{BH}}/10\,\Msun)^{-0.40}$.
Then, the accretion rate and the luminosity are estimated as $\dot{M}_{\rm{burst}}\sim6\times10^{18}\,{\rm{g\,s^{-1}}}\,(\Sigma/6\times10^2{\,\rm{g\,cm^{-2}}})(R_{\rm{d}}/10^{11}\,{\rm{cm}})^2(t_{\rm{vis}}/3\times10^6\,\rm{s})^{-1}$ and $L_{\rm{burst}}\sim5\times10^{38}\,{\rm{erg\,s^{-1}}}\,(\dot{M}_{\rm{burst}}/6\times10^{18}\,{\rm{g\,s^{-1}}})\sim0.4\,L_{\rm{Edd}}$ for $M_{\rm{BH}}=10\,\Msun$, where $L_{\rm{Edd}}=\eta\dot{M}_{\rm{Edd}}c^2$ is the Eddington luminosity.
The above estimated values of the rise and decay timescales and luminosity agree with the observed values of $t_{\rm{rise}}\sim10\,\rm{days}$, $t_{\rm{decay}}\sim30\,\rm{days}$, and $L_{\rm{burst}}\sim0.1\,L_{\rm{Edd}}$ \citep{1997ApJ...491..312C,2015ApJ...805...87Y}.

\if{The observables of X-ray novae such as the rise timescale, radiated energy, and decay timescale are also reproduced by our isolated BH model.
These observables are evaluated as functions of the mass accretion rate and the disk radius \citep{1998MNRAS.293L..42K,2001NewAR..45..449L}.
These values are comparable to those of the BHs in binary systems.
Therefore, isolated BHs in molecular clouds can potentially produce outbursts with the same rise and decay timescale and radiated energy with the observed X-ray novae.
}\fi

Finally, we estimate the event rate and show that the rate is consistent with the observations.
We assume that, in an outburst phase, the BHs have the same luminosity and duration as the observed X-ray novae of $L_{\rm{burst}}\simeq0.1L_{\rm{Edd}}$ ($\dot{M}_{\rm{burst}}\sim0.1\dot{M}_{\rm{Edd}}$) and $t_{\rm{burst}}\sim{t_{\rm{rise}}+t_{\rm{decay}}}\sim40\rm{\,days}$ \citep{1997ApJ...491..312C,2015ApJ...805...87Y}.
To power the outbursts, BHs with $\dot{M}\simeq3\times10^{-3}\dot{M}_{\rm{Edd}}$ should continue to accrete mass with a duration of
\begin{eqnarray}
t_{\rm{quiet}}&\sim&\frac{t_{\rm{burst}}\dot{M}_{\rm{burst}}}{\dot{M}},
   \label{duty cycle}\\
&\simeq&1300\,\biggl(\frac{t_{\rm{burst}}}{40\,\rm{d}}\biggl)\biggl(\frac{\dot{M}_{\rm{burst}}}{0.1\,\dot{M}_{\rm{Edd}}}\biggl)\biggl(\frac{\dot{M}}{3\times10^{-3}\,\dot{M}_{\rm{Edd}}}\biggl)^{-1}\,{\rm{days}}.
\end{eqnarray}
It should be noted that the derived duration is less than the dynamical timescale of a BH to cross the Bondi radius $t_{\rm{dyn}}$ for $V=10\,\rm{km\,s^{-1}}$, which corresponds to the angular momentum flip timescale of the disk.
Then, the event rate is estimated as
\begin{eqnarray}
\frac{N_{\rm{BH}}}{t_{\rm{quiet}}}\simeq5.5\,\biggl(\frac{N_{\rm{BH}}}{20}\biggl)\biggl(\frac{t_{\rm{burst}}}{40\,\rm{d}}\biggl)^{-1}\biggl(\frac{\dot{M}_{\rm{burst}}}{0.1\,\dot{M}_{\rm{Edd}}}\biggl)^{-1}\biggl(\frac{\dot{M}}{3\times10^{-3}\,\dot{M}_{\rm{Edd}}}\biggl)\,\rm{yr^{-1}}.
   \label{event rate}
\end{eqnarray}
Interestingly, this value is comparable to the rate of the observed X-ray novae \citep[$\sim$ a few $\rm{yr}^{-1}$,][]{1997ApJ...491..312C,2016A&A...587A..61C}. 
Therefore, a part of X-ray novae may be produced by isolated BHs.
In order to study this possibility, we need deep and multi-wavelength follow-up observations to confirm whether X-ray novae occur in molecular clouds, and whether there are companion stars or not. 

\section{Discussion}\label{Discussion}
We discuss the observational strategy to confirm whether an X-ray nova is produced by an isolated BH or not.
The simplest way is to exclude companion (or secondary) stars by follow-up observation.
Typically, secondary stars of low mass BH X-ray binaries are K- or M-type dwarves \citep{2006ARA&A..44...49R}.
Their effective temperature and radius are about $T_{\rm{eff}}\sim4000\,\rm{K}$ and $R_*\sim\,0.5\,R_{\odot}$ \citep{2010A&ARv..18...67T}.

Then, we estimate the apparent magnitudes for optical $V$- ($\lambda_{V}=0.545\,\rm{\mu{m}}$) and near infrared $J$- ($\lambda_{J}=1.215\,\rm{\mu{m}}$), $H$- ($\lambda_{H}=1.654\,\rm{\mu{m}}$), and $K_s$-bands ($\lambda_{K_s}=2.157\,\rm{\mu{m}}$) of the companion stars taking the extinction into account.
We set the distance to an isolated BH $d\sim4\,\rm{kpc}$, which is a typical distance to dynamically confirmed BHs \citep{2016A&A...587A..61C}.
For this distance, the hydrogen column density contributed from the interstellar space, amounts to $N_{\rm{H}}\sim1.2\times10^{22}(n/1\,{\rm{cm^{-3}}})(d/4\,{\rm{kpc}})\,\rm{cm^{-2}}$, where $n$ denotes the number density of the interstellar space.
On the other hand, the column density of the molecular clouds where the BH resides, is $N_{\rm{H}}\sim0.9\times10^{22}(n/10^2\,{\rm{cm^{-3}}})(l/{\rm{30\,pc}})\,\rm{cm^{-2}}$, where we use a typical molecular cloud size of $l\sim30\,\rm{pc}$ \citep{2017ApJ...834...57M}.
We see that these contributions are comparable, and we set the total column density as $N_{\rm{H}}=2\times10^{22}\,\rm{cm^{-2}}$ to estimate the extinction.
For the column density, the extinction of each band results in $A_{V}\simeq12\,\rm{mag}$, $A_{J}\simeq3.2\,\rm{mag}$, $A_{H}\simeq2.0\,\rm{mag}$ and $A_{K_s}\simeq1.4\,\rm{mag}$, where we use $A_\lambda/N_{\rm{H}}\simeq6.0$, $1.6$, $1.0$, and $0.7\times\rm{10^{-22}\,cm^{2}\,mag}$ for $V$-, $J$-, $H$-, and $K_s$-bands, respectively \citep{2003ARA&A..41..241D}.
Then, the apparent magnitudes in these bands become $V\simeq33.6\,\rm{mag}$, $J\simeq23.4\,\rm{mag}$, $H\simeq22.3\,\rm{mag}$ and $K_s\simeq21.9\,\rm{mag}$.
We find that optical follow-up observations are extremely difficult to detect secondary stars.

\begin{figure}
\includegraphics[width=0.7\columnwidth,angle=270]{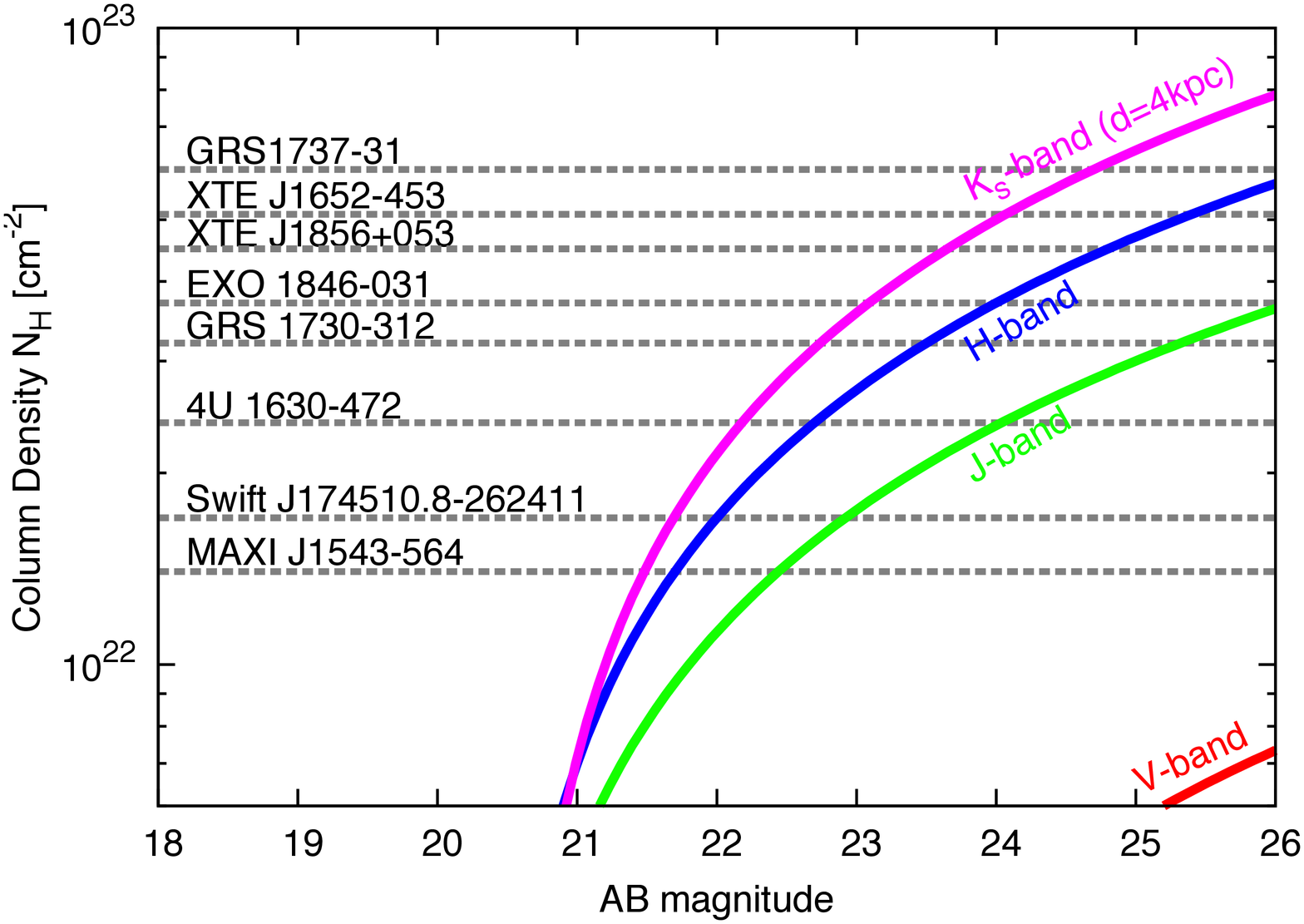}
\caption{The required depth to detect the companion stars of the observed X-ray novae with the measured column densities.
Red, green, blue, and magenta curves show the apparent magnitude of the companion stars, when the companions whose temperature and photospheric radius of $T_{\rm{eff}}\sim4000\,\rm{K}$ and $R_*\sim\,0.5\,R_{\odot}$, are located at $4\,\rm{kpc}$ from Sun. 
The horizontal dotted lines represent the measured column densities for X-ray novae which are not confirmed to have companions.
}
\label{fig mag}
\end{figure}

We discuss an observational strategy to confirm the absence of companions in near infrared bands.
When an X-ray nova is produced by a binary system, we can find the secondary star by deep near infrared follow-ups.
In the actual observations, we can use rough estimations of the column density by using the absorption signature of the soft X-ray spectrum \citep[][and references therein]{2007A&A...469..807L,2016A&A...587A..61C}, in order to evaluate the necessary depths for follow-ups.
In Fig. \ref{fig mag}, we show the required depth to detect companion stars if they exists.
Some observed X-ray novae whose column densities are evaluated by fitting of the absorbed soft X-ray spectrum \citep[][and references therein]{2016A&A...587A..61C} are also shown.
The selected X-ray novae are located at the low galactic latitude position $b\lesssim1.5\,\rm{deg}$ because their host molecular clouds have small scale heights (see Table \ref{table ism}).
Horizontal dotted lines represent the column density of each transient.
Red, green, blue, and magenta curves show the estimated apparent AB magnitude, when the companions are located at $4\,\rm{kpc}$ from Sun, including extinction in $V-$, $J-$, $H-$, and $K_s$-bands, respectively.
For example, \textit{MAXI J1543-564} has a column density of $\simeq1.4\times10^{22}\,\rm{cm^{-2}}$.
With this density, we should conduct follow-up observations as deep as $\sim22.5$, $21.6$, and $21.4\,\rm{mag}$ in $J$-, $H$-, $K_s$-bands.
If we can not detect any sources, the transient could be launched by an isolated BH.

It should be also noted that a detection of near infrared sources does not necessarily mean the existence of the companion.
We also detect the disk as a near infrared source.
In this case, we should conduct spectroscopic observations.
When the near infrared source is not the companion but the disk emission of an isolated BH, we do not detect any periodic variation of emission or absorption lines \citep[e.g., $\rm{H}_{I}$ and $\rm{He}_{II}$ emission lines, or neutral metals and molecular absorption lines, as actually observed in BH binary systems][]{2010ApJ...716.1105K}.
In a catalog of BH X-ray novae \citep{2016A&A...587A..61C}, such near infrared candidate sources without periodic variabilities have been already reported, e.g. \textit{IGR J17454-2919}, \textit{XTE J1908-094}, and \textit{SAX J1711.6-3808}.
Among these X-ray novae, \textit{IGR J17454-2919} \citep{2015ApJ...808...34P} and \textit{SAX J1711.6-3808} \citep{2014ApJ...788..184W} are located at the same positions where the near infrared sources have already been detected in the past survey observations, although the positions are crowded with stars and the chance of coincidence is high.
Further follow-up observations are not conducted.
For the other X-ray nova, \textit{XTE J1908-094} \citep{2006MNRAS.365.1387C}, only photometric observations were conducted, and any detections of periodic variabilities were not reported.
We should conduct deeper and more careful photometric and spectroscopic observations for these objects.

Radio follow-up observations are also important to study whether the transients are accompanied by molecular clouds or not.
For example, the X-ray source \textit{1E 1740.7-2942} in the Galactic center was first proposed to be in a giant molecular cloud by radio observations \citep{1991Natur.353..234B,1991A&A...251L..43M}, but later shown to be located behind the cloud by detailed X-ray spectroscopic observations \citep{1996ApJ...464L..71C}.

We also discuss the feedback effect of isolated BHs on the event rate.
Theoretical studies suggests that ADAF solutions allow outflows from the disk systems \citep{1999MNRAS.303L...1B,2004MNRAS.349...68B,2015ApJ...804..101Y}.
Even a small fraction of accretion mass is blown away, the outflow affects the medium around the Bondi radius \citep{2017MNRAS.470.3332I}.
When the feedback works, the accretion rate is decreased, which also makes the outflows weak.
\cite{2017MNRAS.470.3332I} estimated the duty cycle of the self-regulation is about $\sim1/10$.
In this case, the event rate of X-ray novae produced by isolates BHs will be decreased by $\sim1/10$.

When an X-ray flux from the inner disk part is large, it may affect the disk structure \citep{1996ApJ...464L.139V}.
The X-rays ionize hydrogen and the Saha equation \eqref{saha} does not give a correct degree of ionization.
\cite{1999MNRAS.303..139D} studied the effect of the X-ray irradiation and concluded that the disk structure nor the S-curve do not change so much in the quiescent state.
However, if the accretion disk is warped or there is a X-ray irradiation source above the disk plane, the X-ray ionization may affect the hot branch in the S-curve.
This is an interesting future work.

Finally, we remark the difference between our work and \cite{2002MNRAS.334..553A} who also discussed a probability that isolated BHs launches X-ray novae.
They calculated the event rate of X-ray novae produced by isolated BHs by using the mass accretion distribution.
Their calculation predicted much more event rate than the observed one, and they concluded that X-ray novae caused by isolated BHs are unlikely.
However, they did not consider the disk structure and the instability condition, and extremely underestimate the minimum accretion rate needed to produce X-ray novae.
This is why \cite{2002MNRAS.334..553A} overestimated the event rate of X-ray novae powered by isolated BHs.

\section*{Acknowledgements}
We are thankful Aya Bamba and Shota Kisaka for fruitful discussions and comments.
We would like to thank Hitoshi Negoro for carefully reading the manuscript and giving us useful comments.
We also thank Kengo Tachihara, Rei Enokiya, Shu Masuda and Shogo Kobayashi for kindly helping us to check the CO observation map.
TM is especially thankful to Takashi Hosokawa for helpful discussions and daily encouragements.
This work is supported by Grant-in-Aid for JSPS Research Fellow 17J09895 (TM), KAKENHI 24000004, 24103006, 26247042, 26287051, 17H01126, 17H06131, 17H06357, and 17H06362 (KI).








\bsp	
\label{lastpage}
\end{document}